# Histopathology Based AI Model Predicts Anti-Angiogenic Therapy Response in Renal Cancer Clinical Trial

Jay Jasti[1], Hua Zhong[1,2], Vandana Panwar[2], Vipul Jarmale[1], Jeffrey Miyata[3], Deyssy Carrillo[3], Alana Christie[3,4], Dinesh Rakheja[2], Zora Modrusan[5], Edward Ernest Kadel III[5], Niha Beig[5], Mahrukh Huseni[5], James Brugarolas[3,6], Payal Kapur[2,3,*], Satwik Rajaram[1,2,3,*]

*Affiliations:*

1. Lyda Hill Department of Bioinformatics, University of Texas Southwestern Medical Center, Dallas, TX, USA
2. Department of Pathology, University of Texas Southwestern Medical Center, Dallas, TX, USA
3. Kidney Cancer Program, Simmons Comprehensive Cancer Center, University of Texas Southwestern Medical Center, Dallas, TX, USA
4. O'Donnell School of Public Health, The University of Texas Southwestern Medical Center, Dallas, Texas.
5. Genentech, South San Francisco, CA, USA
6. Department of Internal Medicine (Hematology-Oncology), University of Texas Southwestern Medical Center, Dallas, TX, USA
*. Corresponding

## Abstract

Background: Predictive biomarkers of treatment response are lacking for metastatic clearcell renal cell carcinoma (ccRCC), a tumor type that is treated with angiogenesis inhibitors, immune checkpoint inhibitors, mTOR inhibitors and a HIF2 inhibitor. The Angioscore, an RNA-based quantification of angiogenesis, is arguably the best candidate to predict anti-angiogenic (AA) response. However, the clinical adoption of transcriptomic assays faces several challenges including standardization, time delay, and high cost. Further, ccRCC tumors are highly heterogenous, and sampling multiple areas for sequencing is impractical.

Approach: Here we present a novel deep learning (DL) approach to predict the Angioscore from ubiquitous histopathology slides. In order to overcome the lack of interpretability, one of the biggest limitations of typical DL models, our model produces a visual vascular network which is the basis of the model's prediction. To test its reliability, we applied this model to multiple cohorts including a clinical trial dataset.

Results: Our model accurately predicts the RNA-based Angioscore on multiple independent cohorts (spearman correlations of 0.77 and 0.73). Further, the predictions help unravel meaningful biology such as association of angiogenesis with grade, stage, and driver mutation status. Finally, we find our model is able to predict response to AA therapy, in both a real-world cohort and the IMmotion150 clinical trial. The predictive power of our model vastly exceeds that of CD31, a marker of vasculature, and nearly rivals the performance (c-index 0.66 vs 0.67) of the ground truth RNA-based Angioscore at a fraction of the cost.

Conclusion: By providing a robust yet interpretable prediction of the Angioscore from histopathology slides alone, our approach offers insights into angiogenesis biology and AA treatment response.

## Introduction:

Patients with metastatic clear cell renal cell carcinoma (ccRCC) are treated with anti-angiogenic (AA) therapies (e.g., vascular endothelial growth factor tyrosine kinase inhibitors VEGF-TKIs), immune checkpoint inhibitors (ICI), mammalian target of rapamycin (mTOR) inhibitors and a hypoxia inducible factor (HIF)-2 inhibitor, either in combination or as monotherapy (1). However, individually, these therapies give rise to heterogenous outcomes, and none uniformly benefit all patients. Further, therapeutic decisions between regimens combining ICI and AA are not evidence-based, and it is unclear whether this combination is truly synergistic (2). In fact, recent biomarker analyses suggest that some ccRCCs are exclusively responsive to one or the other (3-5). Thus, there is a critical need for predictive biomarkers of treatment response.

Multiple strategies have been deployed to develop biomarkers for ccRCC, however none have advanced to the clinic. Arguably, to date, the most promising approaches have used RNA sequencing (RNAseq). McDermott et al. (5) showed that patients with higher expression of angiogenesis related genes (Angioscore) exhibited better response to AA therapy in the phase 2 IMmotion 150 trial. Similar analyses were performed in the phase 3 IMmotion 151 and Javelin trials (4, 6). Notably, an RNA-based biomarker is now being explored prospectively in the OPTIC RCC trial (NCT05361720) (7).

However, transcriptomic-based biomarkers are challenging for everyday clinical use. Not only are they time-consuming but also suffer from experimental variability as readouts are sensitive to sample quality and batch effects (8-10). Moreover, given the high cost, typically only a small portion of the tumor is profiled, which may be problematic in notoriously heterogenous ccRCC (11-13).

Some of these limitations may be addressed using Hematoxylin and Eosin (H&E) stained histopathologic slides which are generally available from multiple representative areas of the tumor, are economical, and are ubiquitous for clinical diagnosis and prognostication. Importantly, the Angioscore is largely based on genes expressed by endothelial cells, which are visually distinct in H&E slides and can be segmented using computational models in both ccRCC (14) and other cancers (15). Interestingly, while immunohistochemical (IHC) staining of CD31 (marker for endothelial cells) correlates with the Angioscore, the relationship was noisy, possibly due to technical challenges associated with quantifying IHC staining (5). Thus far, the approaches that have shown most promise in predicting gene expression signatures from H&E slides have been deep learning (DL) models that directly make these predictions without reference to underlying cell types (16-20). While these models have promising predictive power, their lack of interpretability poses a challenge.

Here, we present a visually interpretable DL biomarker for histopathological slide image analysis that correlates with the RNA-based Angioscore as a means to infer response to AA therapy in ccRCC. A key feature is the use of both the RNA Angioscore and physical endothelial cells as training ground truth, which increases robustness and provides interpretability. Specifically, the model generates a visual representation of the vascular network that provides the basis for the predictions. We show that this predicted Angioscore from H&E images alone (H&E DL Angioscore) correlates strongly with the RNA based Angioscore in multiple independent cohorts including from the IMmotion 150 clinical trial. Further, we explore the relationship of the H&E based Angioscore with various clinical and prognostic variables including grade, stage, and gene status. Finally, we validate the performance of the model as a predictor of response to AA therapy on a real-world clinical data set and on data from the IMmotion 150 clinical trial.

## Results:

**Building a H&E Based Deep Learning Model to Predict the RNA-based Angioscore.**

To predict the Angioscore and evaluate its ability to predict response to AA therapy, we developed a workflow that applies a DL model to tumor regions identified in whole slide images (WSI) of H&E-stained ccRCC slides (Fig. 1A, Methods). A key design principle in developing this workflow was that the output of the model be visually interpretable. To achieve this, we trained the model to predict a vascular network ("vascular mask" based on CD31 IHC). The final output of the model – termed as the H&E DL Angioscore was intended to match the Angioscore from RNA while being a simple summary of a visually interpretable prediction: the "vascular mask" i.e. the network of endothelial cells in CD31 IHC (Fig. 1B). This approach makes it possible to visually interpret the basis of model predictions, allowing us to build confidence in its performance and diagnose deviations from expected trends (Fig. 1C).

To overcome the limitation that we did not simultaneously have RNA and CD31 IHC on the same samples, we built a mixed DL model that separately predicts the vascular mask and the RNA Angioscore and enforces consistency between these two predictions (Fig. 1B-C, Supp. Fig. 1, Methods). To build the CD31 model (i.e. vascular mask prediction), H&E and IHC images were computationally aligned, and pixels were given ground truth assignments as CD31 positive or negative based on an IHC model (Supp. Fig. 2, Methods: UTSW CD31 Re-stain dataset). A U-Net (21) model with a Resnet (22) backbone was then trained to recover these ground truth positive/negative pixel assignments based purely on the H&E input. The RNA Angioscore prediction model shares the encoder portion with the vascular mask model but has its own subnetwork that predicts a single Angioscore value for each patch. This model was trained by using public data from the TCGA KIRC: for each slide image patches from the tumor regions served as input, and the matching RNA Angioscore (23) served as target ground truth. Finally, to establish consistency between the two arms, we required concordance between the predicted RNA Angioscore and the percentage of positive pixels from the CD31 prediction. To train our models we alternate between patches of data with CD31 and RNA ground truth, updating the network weights to maximize agreement with the appropriate ground truth as well as consistency between the Angioscore and CD31 predictions.

**Validation of H&E DL Angioscore Model**

We first tested the performance of our DL model on held out portions of our training sets. We compared the endothelial cell outputs of the CD31 arm to the CD31 IHC (Supp. Fig. 2) and found good segmentation performance with a tendency to overpredict the boundaries of the CD31 mask (Methods, precision=0.53, recall=0.66, F1=0.58, and Supp Fig 3). In addition, an expert genitourinary pathologist (PK) reviewed the output masks in conjunction with the H&E images in the TCGA cohort to ensure there were no systematic over or under predictions leading to deviations with the ground truth RNA Angioscore. Next, we tested the performance of our RNA Angioscore. In principle, our model has two readouts that should correlate with the RNA Angioscore: the output from the RNA score arm and the % of positive pixels from the CD31 mask arm. We first confirmed that, although connected by a non-linear transform, these two readouts are strongly correlated (Supp. Fig. 4 right column, Spearman correlation = 0.95). Since results would be unchanged by choice of readout, to maximize interpretability we use the CD31 mask arm for all further analyses. We henceforth refer to the % positive pixels from this arm as the H&E DL Angioscore. Next, we compared the H&E DL Angioscore to the RNA Angioscore on the held-out portion of the training TCGA cohort (consisting of ~33% of the slides) and found a correlation of 0.68

(Fig. 2A). Notably, by comparing models that are trained based on a single type of data, either only RNA or CD31 IHC, we found that performance was improved by addition of our mixed model approach (Supp. Table 1).

**Performance of H&E DL Angioscore Model in predicting RNA based Angioscore on Real-world and clinical trial datasets.**

Next, we tested the ability of our model's H&E DL Angioscore to predict the true RNA Angioscore on two previously unseen independent cohorts (UTSeq and IMmotion 150). Unlike the TCGA, where the RNA and H&E tissue are extracted from different samples, and hence potentially impacted by intra-tumor heterogeneity, the tissue for H&E and RNA seq analysis are spatially matched in these two cohorts. We first tested the performance on the UT Sequencing cohort (UTSeq, Methods), which is a custom morphology guided sequencing dataset with 196 samples. This UTSeq cohort has a tight alignment between morphology and sequencing as punched areas for sequencing analyses were confirmed to be flanked by matching H&E images on top and bottom. In this cohort, we observed a correlation of 0.77 between the RNA Angioscore and our predicted H&E DL Angioscore (Fig. 2B, Supp. Fig 5). Next, we tested the performance of our model on the IMmotion 150 clinical trial slides (Fig. 2C). This cohort contains 227 available H&E slides from different patients, and the RNA was extracted by macro-dissection of serial sections (5). We obtained similar strong agreement (correlation of 0.73) between the H&E DL Angioscore and the RNA-based score for the IMmotion 150 cohort. Importantly, our model performs well (Supp. Fig 6) regardless of tissue extraction site (primary vs metastatic) or procedure (resection vs biopsy). In both the IMmotion 150 (Fig. 2D, correlation of 0.73 vs 0.65) and the subset of UTSeq samples with both RNAseq and CD31 IHC (Supp. Fig. 7, correlation of 0.61 vs 0.45), the H&E DL Angio correlates much better with RNA Angio than the CD31 IHC does.

**Exploring H&E DL Model Angiogenesis prediction for biomarker discovery**

Having validated our H&E based model's quantification of angiogenesis, we asked how its predictions correlated with well-established prognostic variables. We leveraged our previously published Tissue Microarray (TMA) cohort with over 800 punches that otherwise lacks gene expression data (24) (Methods). We found an inverse correlation between the World Health Organization/ International Society of Urological Pathology (WHO/ISUP) nucleolar grade and the H&E DL Angioscore (Fig. 3A). We extended these analyses to the cohorts where we have RNA and observed a similar inverse relationship between grade and Angioscores based on RNA or H&E (Supp Fig. 8). But interestingly, the H&E DL Angioscore for a given grade is more consistent across cohorts than from RNA, suggesting our model could provide a means to overcome batch effects that impact transcriptomic signatures and pose a significant challenge for clinical adoption.

Extending our analysis to TNM stage (Fig. 3B), we found that tumors with high stage (stage 3 and 4) are associated with lower angiogenesis than low stage ccRCCs (stage I and II). Similarly other prognostic factors like tumor size (Supp. Fig. 9A) and presence of sarcomatoid features (Supp. Fig.9B) negatively correlate with the H&E DL Angioscore, supporting the hypothesis that angiogenesis is progressively reduced with tumor progression. We recently developed an atlas of ccRCC architectural subtypes by identifying recurrent patterns of vascular structure and demonstrated that some "indolent" architectures were indicative of a favorable prognosis while other more "aggressive" ones correlated with poor clinical outcome(11). In the UTSeq cohort we found the expected (11, 25, 26) reduction of angiogenesis (both on RNA and H&E DL Angioscore) with more aggressive architectural patterns (Supp.

Fig. 10). Overall, our results show that our DL model accurately captures vascular network, and that vascular network is strongly associated with tumor architecture, as well as tumor grade and stage.

We have previously shown that BAP1- and PBRM1-loss drive tumor grade and aggressiveness and ccRCC with PBRM1- loss tend to be of low grade, while BAP1 loss are of high grade. Given the correlation of grade and stage with frequently mutated driver genes in ccRCC (24), we sought to capture the effect of BAP1 and PBRM1 loss on angiogenesis. BAP1 loss significantly correlated with lower H&E DL Angioscore relative to wild-type (4.9 vs 9.4, $p = 3.67 \times 10^{-7}$), whereas PBRM1 loss leads to a slight (9.9 vs 9.4) albeit not-statistically significant increase (Fig. 3C).

Next, we tested the extent to which the H&E DL Angioscore correlated with survival. As might be expected based on our previous results, there is a strong relationship between overall survival and the H&E DL Angioscore with a c-index of 0.75. To visualize this relationship using the Kaplan-Meier method, we stratified patients based on their H&E DL Angioscore. To avoid overfitting on this dataset in determining optimal cutoffs, we turned to the TCGA data, that is representative of the ccRCC spectrum, and evaluated the effect of different choices of cutoffs (Supp. Fig. 11). We observed two peaks in p-value, one that separated out all the highly angiogenic samples with good prognosis and another that sequestered the low Angioscore samples with unfavorable prognosis (Supp. Table 2). These cutoffs also translate to other cohorts, for example, separating out our previously described indolent, intermediate, and aggressive architectural patterns (11) in the UTSeq cohort (Supp. Fig 10A). Based on these observations we applied the low and high peak thresholds from the TCGA to the TMA cohort and performed a three-class stratification (Fig. 3D) which provides HR values of 6.9 (3.7-12.8) and 2.9 (1.7-4.9) for the high and medium DL Angio groups against the low, suggesting that the H&E DL Angioscore can effectively stratify patients with different outcomes.

**H&E DL Angioscore Predicts AA therapy response.**

Given that RNA Angioscore is a predictor of response to AA therapy, we next sought to test the predictive performance of our H&E DL Angioscore. First, we applied our model to a real-world cohort consisting of 145 patients treated at UTSW who received single agent first line AA for metastatic ccRCC between 2006 to 2020. Using Time-to-next treatment (TNT) as a proxy for treatment efficacy we found a c-index of 0.6 with the H&E DL Angioscore. Next, we performed Kaplan-Meir and Cox-proportional hazards analyses, by stratifying the patients based on their H&E DL Angioscore. As this cohort was restricted to metastatic cases (which as expected had few cases with higher Angioscore), we performed a two-class stratification to identify low angiogenic tumors using the threshold described above (Fig. 4A; threshold H&E DL Angioscore of 5.66) and obtained a HR of 0.64 (95% CI: 0.45-0.91) with a p value of 0.012. Interestingly, best stratification with hazard ratio of 0.41 (95% CI: 0.26-0.64) with a p value of $8.7 \times 10^{-5}$ was obtained using an even lower threshold Angioscore of 2.34, likely indicating that this cohort has a greater proportion of aggressive tumors with low angiogenesis that would not benefit from AA therapy.

Finally, as a gold standard test of our H&E model to predict AA therapy response, we analyzed the IMmotion150 clinical trial. This trial contains three treatment arms, a TKI Sunitinib, an ICI atezolizumab, and the combination of atezolizumab with an anti-VEGF antibody, bevacizumab. We compared the predictive value of the H&E DL Angioscore to that of the previously reported RNA Angioscore, as well as

CD31 IHC, for response to sunitinib. First, we examined the relationship to Progression free survival (PFS), and found c-index of 0.66, 0.67, 0.55 for the H&E, RNA and CD31 IHC based assays respectively. Next, we stratified the patients into low/high angiogenesis groups based on the median score for each assay following the original IMmotion150 publication (results are even stronger for H&E DL Angioscore with the TCGA based cutoff above; Methods, Supplementary Table 2). We generated Kaplan Meir curves and performed Cox-proportional hazards calculations (Fig. 4B, Supp. Fig. 12), both demonstrating that the RNA and H&E based predictions of Sunitinib response are comparable and far superior to the CD31 IHC. This point was further reinforced by separate analyses comparing our three assays in terms of: a) the fraction of patients who responded to Sunitinib among high/low Angiogenesis groups (Fig. 4C) and b) the AUC in predicting the objective response (responder or not) based on Angiogenesis (Fig. 4D). Additional information on Sunitinib treatment response and treatment response to other drugs used in IMmotion 150 trial are shown in Figs. Supp. 13 14 respectively. Interestingly, our score (like the RNA based score) captures the previously reported(27, 28) inverse relationship between angiogenesis and response to ICI (Supp. Fig 14, relative heights of bars across arms). Taken together, our results show that DL based model that can predict Angioscore solely from H&E images nearly rivals the gold-standard RNA-based Angioscore and greatly outperforms the CD31 IHC in both real work and clinical trial data (IMmotion150).

**Discussion:**

Anti-angiogenic (AA) drugs are approved for patients with metastatic ccRCC, either as monotherapy or in combination with ICI. Recent data indicate that ccRCC with high levels of vascularity respond better to AA therapy. Indeed, in both IMmotion 150 and IMmotion 151 trials, high expression of a 6-gene Angioscore signature that included *PECAM1* (gene coding CD31) was associated with improved progression free survival (PFS) in patients treated with AA agent- Sunitinib(5, 6) . These results from IMmotion 150 and IMmotion 151 were confirmed by analysis from the JAVELIN renal 101 trial(4). However, clinical adoption of transcriptomic gene signatures has been challenging due to difficulties in standardization, applicability across cohorts, and high cost, particularly given the notorious heterogeneity of ccRCC tumors(27, 28). To address this, we present a robust deep learning (DL) model predicting the Angioscore from H&E-stained slides, offering a scalable and cost-effective alternative to RNA-based assays.

The output of our model was strongly predictive of the RNA Angioscore across multiple cohorts, with Spearman correlation of (0.77/0.73) on both unseen real world and clinical trial datasets (UTSeq/IMmotion150). Our H&E DL Angioscore allowed us to explore the relationship of angiogenesis with various prognostic variables on a large cohort without the need for RNA-seq analysis. Our analysis not only reinforces but provides a platform to operationalize the notion that in ccRCC angiogenesis inversely correlated with tumor aggressiveness. Finally, in the IMmotion 150 clinical trial data, we found that the ability of our model to predict response to Sunitinib rivaled that of the RNA.

Our H&E based DL model was trained using both RNA and CD31 IHC – to maximize interpretability and robustness – and offer advantages over each of these assays. Notably, our H&E DL Angioscore surpasses the predictive capabilities of CD31 IHC, in predicting both the gold standard RNA Angioscore and treatment response. Future studies are needed to determine if this is purely due to technical challenges in performing and quantifying CD31 IHC, or whether the combined training with RNA leads to biologically meaningful differences in the vascular masks. Similarly, our model matched the performance of the gold standard RNA Angioscore in prediction of response. Interestingly, this is true even on biopsies, where the 2D sections provide far less tissue than available for sequencing. Importantly, our model offers a more standardizable measurement across diverse samples and mitigating the impact of RNA batch effects. As H&E slides are far more ubiquitous and affordable than transcriptomic assays, it is feasible to profile multiple areas within heterogeneous ccRCC tumors, and our results make a compelling case for pursuing clinical biomarkers based on applying DL to histopathology slides.

Our model is the first to predict response to anti-angiogenic therapy in ccRCC from H&E images alone. The distinguishing aspect of our DL model is that by training on both RNA and CD31 we can provide visually interpretable predictions, and thereby overcome its limitation as a "black box". This is critical for quality evaluation and clinical adoption. In multiple cancers, previous efforts have demonstrated prediction of specific genes and pathways from H&E images (16-20). Alsaffin (13) et al use a transformer-based DL model to predict bulk RNA sequence scores for about 30,000 genes from kidney cancer. Yet, because these predictions are essentially black boxes, they are hard to decipher. Multiple manuscripts identify immune cells from H&E images, but there are few that connect this quantification directly to response prediction without inclusion of additional assays (29-31). In ccRCC, classic image analyses (as opposed to deep learning) have been able to predict endothelial cells, with lower fidelity (14). The ability of such models to predict the AA response has not been tested. However, endothelial cells may not be enough, as our model outperformed the CD31 IHC in predicting both the RNA Angioscore and ultimately response.

In addition to AA therapy, patients with metastatic clear cell renal cell carcinoma (ccRCC) are often treated with ICIs in combination with AA or as a monotherapy. However, it is unclear whether ICI and AA are truly synergistic (32) and there is evidence suggesting for tumors that respond predominantly to one or the other therapies (3-5). Indeed, our data on the IMmotion 150 trial suggests that although high-Angio patients are more likely than low-Angio patients to respond to Sunitinib, the opposite is true for the arms including ICI.

The ability of our model to predict the Angioscore has applications beyond predicting response. Induction of angiogenesis is considered a hallmark of cancer(33), and is central to ccRCC biology, playing an important role in the definition of various molecular subtypes (6). Our deep learning model enabled an in-depth exploration of biological phenomena, leveraging the vast amounts of archival tissue where we lacked transcriptomic profiling. For instance, it confirmed our previous findings on architectural patterns, demonstrating that in ccRCC, instances with low vasculature exhibit significant associations with high nucleolar grade, TNM stage, and shorter overall survival (11). It also supports findings from our genetically engineered mouse model that found Bap-1-deficient to be less vascular than *Pbrm1*-deficient tumors (34).

While our study showcases the potential of H&E-based biomarkers, improvements are necessary for clinical application. Increasing training data across diverse cohorts will enhance reliability and robustness to variations in staining or microscopes. The current model is only applied to tissue from the tumor regions in a WSI and thus improving region identification models could improve the model's performance. Additionally, although our models perform well on biopsies, it is conceivable that small tissue samples may benefit from serial sections. Finally, the performance of the model needs to be validated on data from additional clinical trials with arms that incorporate AA therapy.

Our approach also lays the groundwork to go beyond re-capitulating the RNA Angioscore. We anticipate future studies will examine the potential to improve response prediction by a) considering the variation of Angioscores (rather than a single average level) across multiple H&E slides from a tumor to account for intra-tumor heterogeneity, and b) directly predicting response from images without using the RNA Angioscore as an intermediate target.  Additionally, multi-modal approaches that combine different modalities such as pathology, radiology and genomics will likely result in a more comprehensive biomarker. We also expect this approach can be refined for predicting responses to AA specific agents or adapted to predict immunotherapy responses.

In summary, our DL model is the first to predict anti-angiogenic therapy response in ccRCC solely from H&E images, offering a cost-effective and interpretable alternative to RNA-based assays. By bridging the gap between molecular insights and clinical feasibility, our work sets the stage for transformative advancements in ccRCC therapeutics.

**Methods:**

**Cohorts:** We made use of several cohorts, each consisting of whole slide images of formalin fixed paraffin embedded (FFPE) H&E slides scanned at either 20X (~0.5 microns per pixel) or 40X (~0.25 microns per pixel). Our model was based on 20X images, so all 40X slides were down sampled by a factor of 2.

1. TCGA KIRC dataset: This is the primary training dataset for the RNA prediction. It was downloaded from NIH GDC Data Portal (23, 35, 36) and consists of 519 whole slide images and associated data (transcriptome, patient-related data such as overall survival). Several slides were excluded from our analysis: 14 exhibited frozen sample-like artifacts, 5 had non-ccRCC-like pathology, 2 had only benign renal parenchyma, 2 slides had imaging/staining artifacts, 6 slides were duplicates from the same patient, and 28 lacked RNA information in the pan-cancer dataset. The remaining set of 462 slides was split 2:1 for model training and validation purposes.  Majority of TCGA images are available in 40X magnification, and a small fraction are available at 20X.
2. UTSW CD31 Re-stain dataset: This is the primary training dataset for the vascular mask prediction. It consists of 17 slides that capture the spectrum of ccRCC morphologies, displaying a range of grades, tissue architectures and vascularity.  These slides were first stained with H&E and imaged at 20X magnification using an Aperio scanner. Next, the slides were de-stained, re-stained with antibody for CD31 (clone JC70A; Agilent CA) and re-imaged. Of the 17 slides, 13 were used for training and 4 were held out for testing.
3. UTSW Multiregional sequencing dataset (UTSeq Data): This dataset was used to validate the H&E DL Angioscore predictions as well as predicted vascular masks (on a small set of samples with CD31 staining). It consists of 161 H&E-stained slide pairs from 27 patients.  Multiple samples (punches) were taken from the same patient and RNA measurements with transcriptomic profiles were obtained. Each FFPE punch had flanking top and the bottom H&E-stained images. Serial sections were taken from a smaller set of 36 samples, and they were stained with CD31 and imaged.  The majority of this dataset was imaged at 40X (H&E), with a smaller fraction was imaged at 20X (CD31 IHC).
4. IMmotion150 Dataset: This dataset was used to validate the Angioscore and TKI response predictions as well as compare them to CD31. From the 305 patients enrolled in the IMMotion150 (5), we had access to 239 whole slide (WS) digital images of H&E-stained slides out of which 13 were excluded due to duplicates (n= 5) or quality issues (significant artifacts resulting in extremely low evaluable tumor fraction). The remaining 226 WS images were used along with the associated information on drug response data, RNA and CD31 information as described previously (5). Most of the study was restricted to patients treated with sunitinib, and we analyzed all patients with available data for each assay type (H&E:69, RNA: 63, CD31: 61).

5. UTSW TKI Response Dataset: This dataset consists of 145 H&E-stained slides taken from patients undergoing Anti-VEGF treatment. The Kidney Cancer Explorer - an IRB-approved data portal at UTSW with clinical, pathological, and experimental genomic data – was queried for any patients who received first line VEGF-I for metastatic renal cell carcinoma between 2006 to 2020 (37). The medications included in the search were axitinib, bevacizumab, cabozantinib, cediranib, pazopanib, sorafenib, sunitinib, and tivozanib. 355 patients were found who met these criteria, out of whom 180 had H&E-stained images already available to us for analysis. Data for patients whose treatment was stopped due to toxicity was excluded and the final dataset consisted of 145 patients.
6. Tissue Microarray (TMA): This dataset is a combination of TMA datasets that were described previously (24). Most patients were represented by multiple TMA punches. Grade information is available at the punch level (811 punches) whereas other information (Tumor size, stage, overall survival, sarcomatoid status) was available at patient level (520 patients). BAP1 and PBRM1 protein status (as assessed by IHC (38)) was available for a subset of 304 patients.

**RNA Angioscore Calculation:** For the UTSEQ data, the raw transcriptome sequencing data was processed by the SCHOOL (39) with human reference genome version GRCh38.86 and Fragments Per Kilobase of transcript per Million mapped reads (FPKM) genes were generated. FPKM was normalized to Transcripts Per Kilobase Million (TPM), then log-transformed with 1 added to avoid taking log of zero. For other cohorts, namely TCGA and IMmotion 150, TPM data was directly obtained. The signature genes for determining Angioscore are VEGFA, KDR, ESM1, PECAM1, ANGPTL4 and CD34 following the IMmotion 150 and IMmotion 151 studies (5, 6). The Angioscore for each tumor sample was computed by the mean log transformed TPM of the Angio signature genes.

**Training Data Generation:**
The H&E DL Angio model simultaneously predicts, from H&E images, a CD31 trained "vascular mask" and the RNA Angioscore. The training data for each of these predictions was generated as follows.

<u>CD31 Training Data:</u> Ground truth data from the UTSW CD31 Re-stain cohort consisted of 416x416px H&E patches with matching "vascular" masks of the same size, with each pixel assigned as CD31 positive or negative as follows:

1. Generation of aligned H&E and CD31 IHC image patches: H&E and CD31 stained slides were registered using a two-step process: an initial manual rigid registration at the slide level performed in QuPath (40) to align the slides, followed by an automatic non-rigid registration at the local image level to align the shared hematoxylin channels of the IHC and H&E images. Non-rigid registration was done using patches of 512x512 pixels with SimpleElastix's multi-resolution, pyramid registration framework (41). Each image pair goes through affine registration first and then deformable registration using B-splines. A smaller 416x416 pair of patches were extracted from the center of the registered image pair to remove any edge effects.
2. Binarization of CD31 IHC: We trained a classifier to identify the CD31 positive areas in IHC-stained images and distinguish them from CD31 negative and non-specific/artifactual DAB staining.

Specifically, we generated IHC images with manually annotated ground truth assignments of CD31+/CD31- /artifact and trained a U-Net based model. We validated the performance on the classifier on 20 similarly constructed IHC image, ground truth mask pairs from 4 slides that were not used in the training of the model (Supp Fig 2) and then used it to generate the ground truth vascular mask for our H&E classifier. This model was then applied to the IHC patches in our CD31 Training data to generate a binary mask with pixels classified as CD31 positive or negative (pixels classified as artifact were treated as exhibiting negative staining).

3. Data Split: 17 slides (with matched H&E and CD31 IHC) were split into 14 for training and 3 for model evaluation. Patches were extracted from tumor regions and registered as described above. In all, 41,918 patches were used in training the CD31 Model, while 5,223 patch pairs were used for evaluation.

RNA Training Data: Ground truth data was extracted from the TCGA KIRC data set. It consists of 416x416px H&E image patches with associated RNA Angioscore (calculated as described above, for the patient from whose slide the patch was extracted).

1. Patch Generation: We first identified tumor regions of slides, with preliminary identification based on a CNN model as described previously (24), followed by manual refinement by an expert pathologist (PK) as needed. We then sought to extract 1200 randomly placed patches within the tumor region of each slide. Note: To prevent oversampling the same areas in slides with limited tumor content, we established a threshold sampling density and in 3% of cases we extracted fewer than 1200 patches to stay within this limit.
2. Data splits: The 462 H&E-stained slides from the TCGA cohort were split 2:1 for model training and validation. This split resulted in 360,999 patches for training and 183,894 patches for validation. Each patch was assigned an RNA Angioscore such that all patches from the same slide have the same Angioscore.

**Model Architecture:** The model essentially consists of three sections (Supplementary Fig. 1). First, the mask prediction arm which takes in an H&E image patch and outputs a predicted vascular mask with two classes (CD31 +/-) has a U-Net architecture with an ImageNet pre-trained ResNet-18 backbone. Second, the Angio score prediction arm, which takes as input an H&E image patch and outputs a single number (the Angio score) shares the encoder arm of the U-Net, followed by several convolutional layers. Finally, the consistency arm, takes the output of the mask arm (namely a binary activation mask), calculates the fractional CD31 positive activation (global average pooling), and performs a learnt non-linear transformation to predict the RNA output from the mask output.

**Model Training:** For any training patch, as only a single type of ground truth (CD31 Mask or RNA) is available, we adopted novel loss functions and training strategies:

1. Batching: The model is trained in batches composed of patches with RNA (TCGA) or CD31 ground truth which have batch sizes of 32 and 4 patches respectively. Each RNA batch of 32 patches is sampled from 4 different slides (thereby allowing us to average the slide level predictions across 8 patches, while also stabilizing the Batch Normalization calculations by sampling multiple slides per batch) while the 4 patches for CD31 are selected randomly.

2. Loss Functions: We use a different combination of loss functions for the two types of batches:

a) RNA Ground Truth: The predicted Angio scores $P^{Angio}$ and $P^{Cons}$ from the Angio and Consistency arms respectively are each averaged across patches from the same slide and compared to the true RNA Angioscore $T^{Angio}$ using a mean square error

$$L_{RNA} = \frac{1}{N} \sum_{S \in B} \left( \left\| T_S^{Angio} - \frac{1}{N_S} \sum_{p \in S} P_p^{Angio} \right\|^2 + \left\| T_S^{Angio} - \frac{1}{N_S} \sum_{p \in S} P_p^{Cons} \right\|^2 \right)$$

Where $S \in B$ indicates the slides $S$ present in the batch, $N = 4$ the total number of slides in the batch, $p \in S$ denotes the patches $p$ in the batch belonging to slide $S$, and $N_S = 8$ denotes the number of patches from slide $S$ in the batch.

b) CD31 Mask Ground Truth: The loss is a sum of i) a segmentation loss $L_{Seg}$ between the true CD31 masks $T^{Mask}$ and the predicted mask $P^{Mask}$ and ii) a consistency loss $L_{Cons}$ comparing the predicted Angio predictions from the RNA and Consistency arms as follows:

$$L_{Seg} = 0.9 \times \text{Dice}(T^{Mask}, P^{Mask}) + 0.1 \times \text{WCCE}(T^{Mask}, P^{Mask})$$
$$L_{Cons} = \text{MSE}(P^{Angio}, P^{Cons})$$

Where Dice denotes the Dice loss, WCCE is a class-weighted categorical-cross entropy and MSE is the mean square error.

3. Augmentation: During training we augmented images using mirroring and color augmentation HED adjust (42) with parameters (0.975,1.025). We also reduced saturation of the input images by a multiplier randomly chosen in the range of [0,0.5] since some of our cohorts have highly faded slides.

4. Training: We first pre-trained the mask prediction arm (i.e. the model without the RNA and Consistency arms) with the segmentation loss only, for 10 epochs with Adam optimizer with a learning rate of $10^{-4}$. The full mixed model was then trained for 10 epochs using stochastic gradient descent optimizer with initial learning rate of $10^{-4}$ and a momentum value of 0.9. Model was stored at every epoch and the best model was selected as the one having the highest Spearman correlation coefficient with the TCGA held out dataset (Supp. Table 3).

**Model Inference:**

a. Stain Normalization: To reduce the impact of slide color variations, prior to model inference, we normalized slides from all cohorts (except the TCGA holdout) to match the color distribution of the training TCGA cohort as outlined previously (24).

b. Patch Generation: Patches were selected randomly from tumor areas as described for the training cohort. In the case of the cohorts where whole slides were profiled (TCGA Holdout, IMmotion150 and UTSW TKI) we targeted 1500 patches per slide as in the training cohort. However, for the UTSeq and TMA cohort where local regions were profiled, we targeted 250 and 1000 patches per region respectively.

c. Sample Level Scores: The model was applied to individual patches from a sample (e.g. a slide or a TMA punch) and the median score across all patches in the slide was reported.

**Ablation models:** We compared our mixed model, which combines a CD31 and Angioscore arm, to the corresponding single arm models:

a. CD31 Model Alone: The segmentation model was trained using only the CD31 mask, H&E image pairs and the model performance was calculated for the TCGA held out data as well as IMmotion 150 data. For each slide, the fraction of pixels called as CD31 positive was correlated with RNA Angioscore and the results are shown in supplementary Table 1.

b. Angioscore Model alone: The Angioscore arm of the model was trained alone using the TCGA training dataset. It has a ResNet-18 based encoder followed by 6 additional convolutional layers. The model was trained for 10 epochs and the best model was selected as the one having highest correlation with TCGA test data.

**Survival Analysis:**

1. Cox-Proportional Hazards: Patients were stratified into groups based on a given threshold level for the H&E DL Angioscore (or RNA/CD31 for IMmotion 150). We chose different Time variables based on the dataset, TCGA: Overall Survival, UTSW TKI Response: Time to Next Treatment, IMMmotion150: Progression Free Survival. A Univariate Cox proportional hazard model was then used to determine the characteristics associated with overall survival. Kaplan Meier curves were generated using the lifelines python package.
2. Optimal H&E DL Angioscore threshold: Overall survival data for TCGA was used and the H&E DL Angioscore threshold was varied and overall survival outcomes for the two groups (above and below the threshold) were calculated. We sought to select the threshold with the lowest p value and high hazard ratio, but found two peaks in the TCGA data, which we then used to stratify patients into Low/Medium/High H&E DL Angioscores. The same thresholds were applied across all other cohorts.

**Data/Code Availability**

H&E images for TCGA KIRC can be downloaded from the TCGA GDC portal, while the corresponding gene expression data is available from cBioPortal. The data for the TMA cohort can be downloaded from *https://doi.org/10.25452/figshare.plus.19324118*. IMmotion150 data, including H&E images, Response, RNA AngioScores and CD31 levels is proprietary to Roche. The anonymized genomic data from 163 patients who granted informed consent to share such data, are made available by Roche at the European Genome-Phenome Archive (EGA) under accession number EGAS00001002928. All other data is available from authors upon reasonable request. The final H&E DL Angio model, and all code used in the manuscript will be released at the Rajaram Lab's public GitHub page upon publication.

**Acknowledgements**

We thank all the patients who provided tissues that enabled this research project. This grant was funded by DOD (KC200285) and in part CPRIT (RP220294) and a Pilot grant from the Lyda Hill Department of Bioinformatics. J.B., A.C., D.R. and P.K. are supported by NIH (Specialized Program in Research Excellence in Kidney Cancer P50 CA196516), J.B., A.C., and P.K. by the Cancer Research & Prevention Institute of Texas (CPRIT; RP180192). J.B. and A.C. are also supported by CPRIT (RP180191). P.K. is supported by NIH (R01CA244579, R01CA154475, and R01DK115986), DOD (KC200294), and CPRIT (RP200233). S.R. is supported by CPRIT (RP220294), DOD (KC200285) and startup funds provided through the Lyda Hill Department of Bioinformatics.

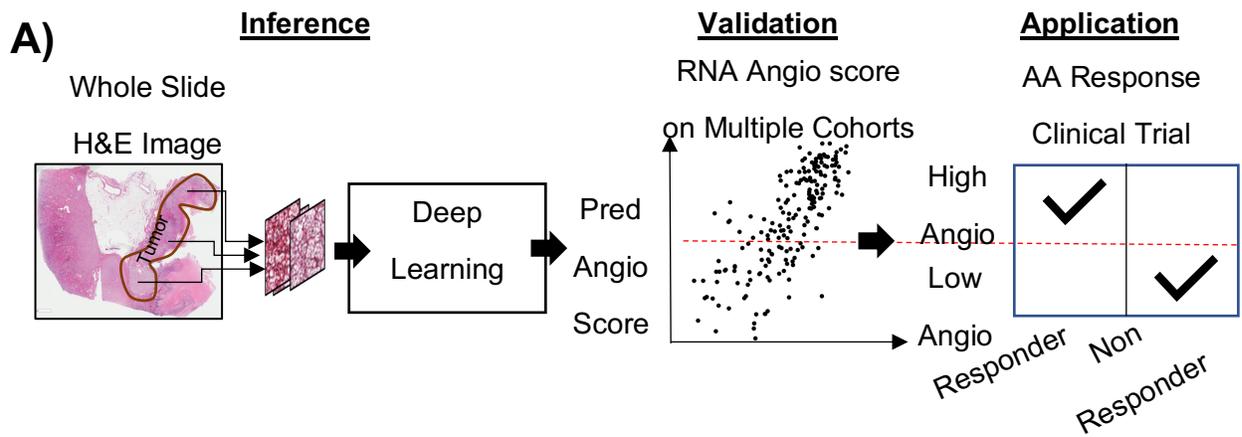

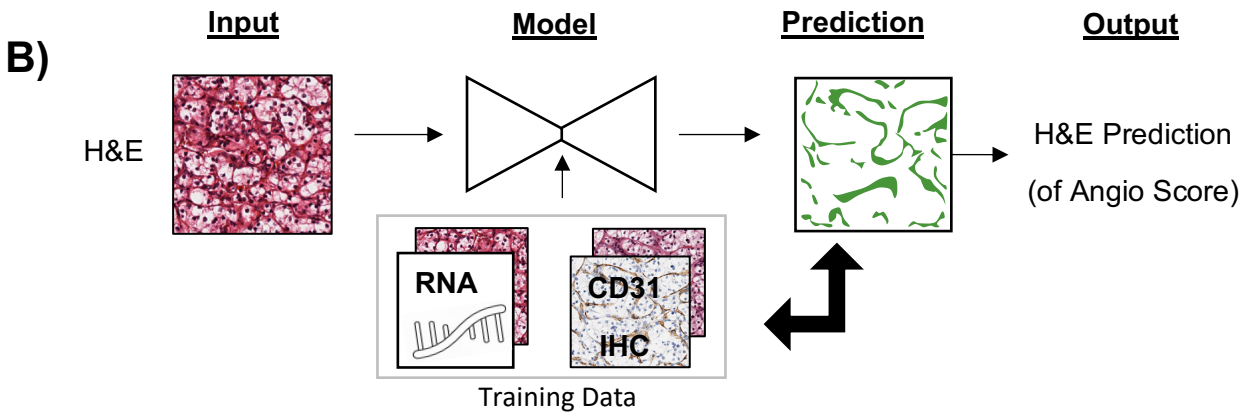

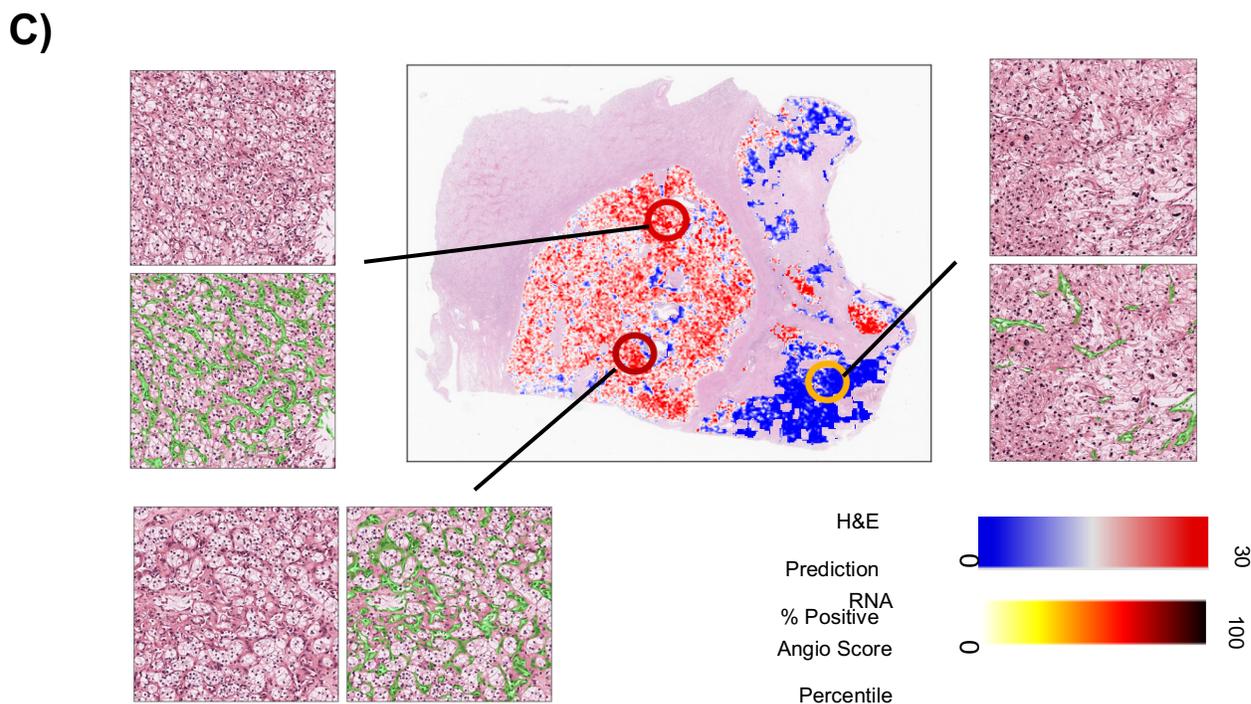

Figure 1 Project and approach overview. A.) Schematic outlining the H&E DL Model development, validation and practical application of the model. The model predicts Angioscores directly from an H&E-stained slide and is validated against the RNA-based Angioscore using multiple independent datasets. The validated model is applied to independent, previously unseen, clinical datasets where its predicted Angioscore is correlated with response to antiangiogenic (AA) therapy. B.) H&E DL Model is an interpretable machine learning model to predict Angioscore from H&E images. Given an input H&E image, the model predicts a vascular mask (green), and the proportion of positive pixels is the output H&E-based Angioscore. Training data consists of two datasets with H&E images matched with RNA-based Angioscores and CD31 IHC (basis of the vascular mask), providing the target ground truth. The model is trained to predict the vascular mask matching the CD31 and the RNA-based Angioscore (see Supplementary Fig 1 and methods for details). C.) Illustration of the model output on a ccRCC case with intra-slide heterogeneity and available multiregional RNA sequencing data. The central plot shows the model H&E Angioscore (blue-white-red colormap) applied to all tumor areas on the slide, calculated based on the local average of the percentage positive vascular mask prediction. For three areas (circles) we also had the ground truth RNA-based Angioscores (circle colors in yellow-red colormap). The zoomed-in H&E images and vascular masks of these areas demonstrate the qualitative agreements in the amount of vasculature and the H&E- based and RNA-based Angioscores.

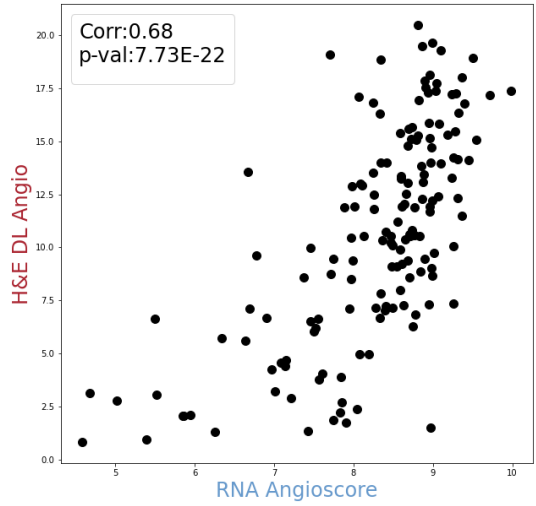
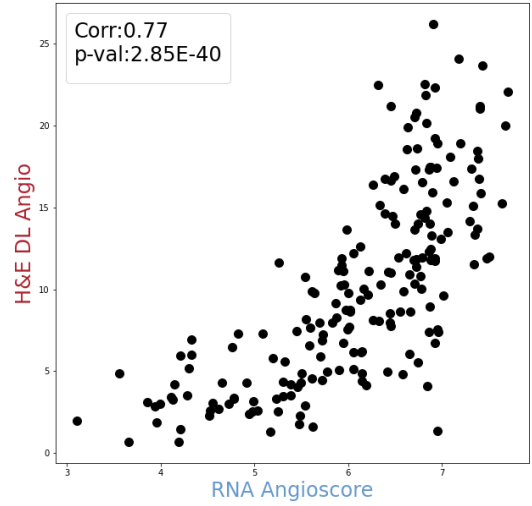
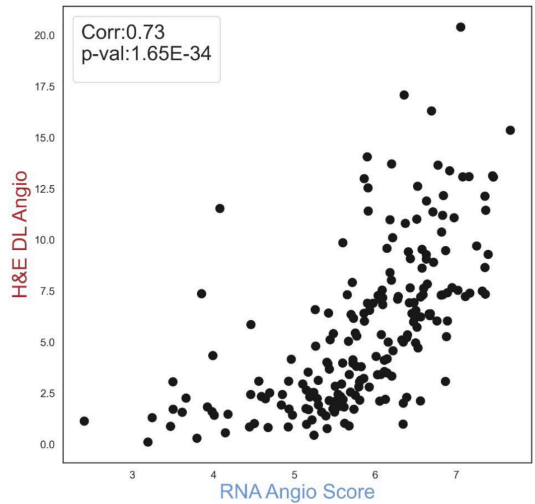
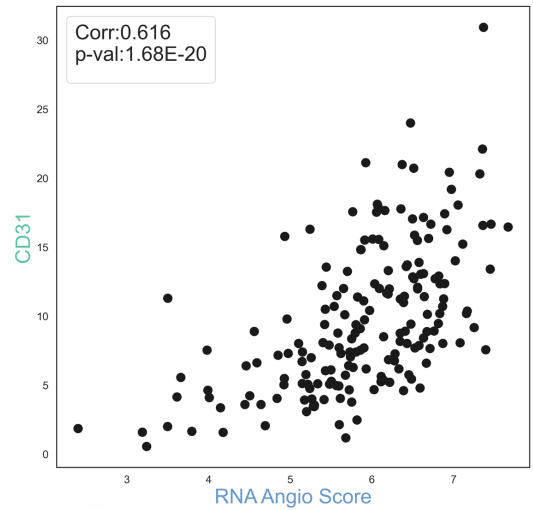

Figure 2 H&E DL Model reliably predicts RNA-based Angioscore across multiple cohorts. Each panel shows a scatter plot (each point represents a sample) comparing RNA-based Angioscore (x-axis) and predicted scores (y-axis). Spearman correlation coefficient along with the p-values are displayed in the legend. A.) Model Performance on TCGA held-out dataset. B.) Model performance on independent UTSeq data set. Predictions were averaged in cases where 2 slides representing the top and flip sides of the block were available. C.) Model performance on IMmotion 150 dataset. D.) Correlation between CD31 measurements and Angioscore measurements for the IMmotion 150 dataset shows that model predictions (i.e., panel C) correlate better with RNA-based Angioscore and outperform CD31 measurements

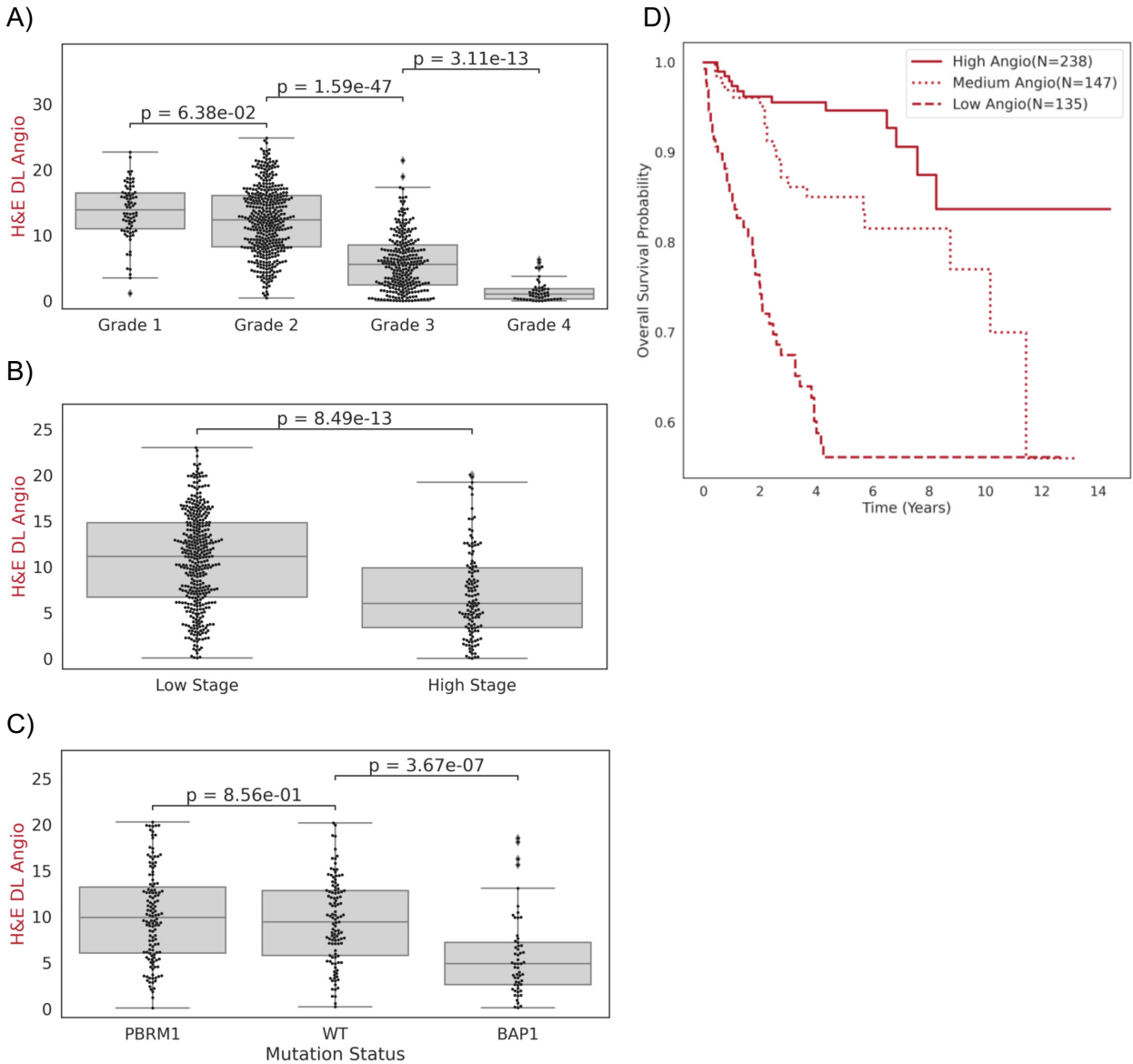

Figure 3 Model predictions correlate with known prognostic variables in independent datasets. The H&E DL Angioscore model was applied to independent tissue microarrays and its output was compared to various prognostic variables including A.) Nuclear Grade, B.) Patient Stage and C.) Functional status of driver genes BAP1 and PBRM1 (the few cases with loss of both BAP1 and PBRM1 are considered as exhibiting BAP1 loss) D.) Kaplan-Meier curves showing overall survival of 520 patients stratified by H&E DL Angioscore (c-index=0.75). Threshold scores for stratification were independently determined from TCGA overall survival dataset as 10.3 and 5.6. Note: Wilcoxon Rank sum tests were used to calculate the p-values. High-Low pair: HR:6.86(3.69-12.75), pVal=1e-09. Medium-Low pair: HR: 2.93(1.74-4.95), pVal: 5.59e-5

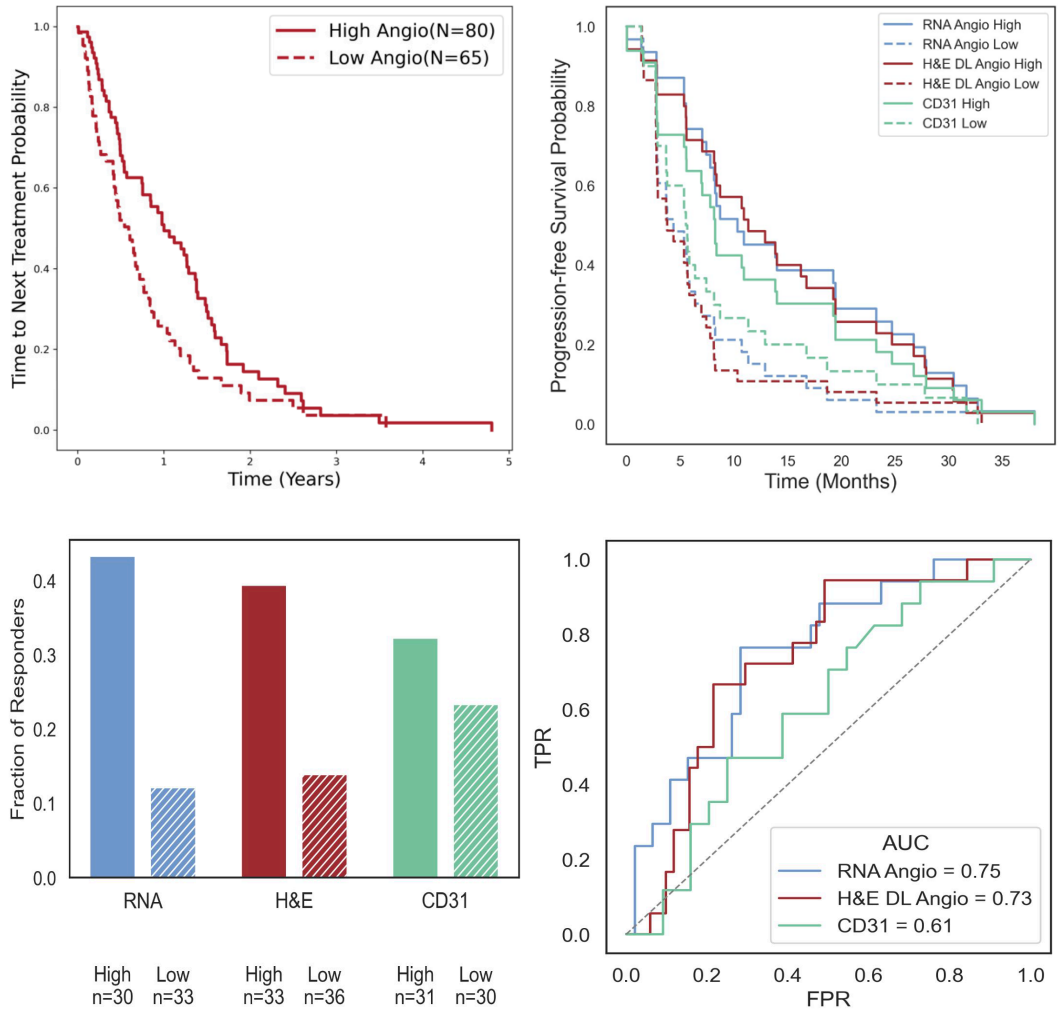

Figure 4 H&E DL Angio Score predicts response to Anti-Angiogenic therapy. A.) Kaplan-Meier analysis of H&E DL Angioscore-based stratification of the AA response of patients from the UTSW-Clinical cohort. Time to next treatment (TNT, x-axis) was used as a measure of treatment response, and patients were stratified into Low and High DL Angioscore based on the threshold (5.6) determined in TCGA prognosis to perform a Cox-Proportional hazards analysis (c-index=0.6, Hazard ratio (Hr): 0.64 (95% confidence interval (CI): 0.45-0.91), p value: 1.23e-2). B.) Kaplan-Meier analysis comparing RNA Angioscore, H&E DL Angioscore and CD31 in their ability to stratify the Sunitinib response of patients from the IMmotion150 clinical trial. Progression free survival (x-axis) was used as a measure of treatment response. Patients were stratified into Low and High groups based on the median levels of their corresponding H&E DL Angioscore and RNA/CD31 Angioscores as in the IMmotion 150 trial. C.) Comparison of the fraction of patients who responded to Sunitinib treatment among the low and high angiogenesis groups as determined by RNA Angioscore, H&E DL Angioscore and CD31 D.) AUC curves comparing how the RNA Angioscore, H&E DL Angioscore and CD31 can distinguish Sunitinib responders and non-responders in the IMmotion 150 clinical trial. Responders were patients having complete or partial response, and non-responders are those with stable or progressive disease.

**References**

1. Motzer RJ, Jonasch E, Agarwal N, Alva A, Baine M, Beckermann K, et al. Kidney Cancer, Version 3.2022, NCCN Clinical Practice Guidelines in Oncology. J Natl Compr Canc Netw. 2022;20(1):71-90.
2. Plana D, Palmer AC, Sorger PK. Independent Drug Action in Combination Therapy: Implications for Precision Oncology. Cancer Discov. 2022;12(3):606-24.
3. Motzer RJ, Tannir NM, McDermott DF, Aren Frontera O, Melichar B, Choueiri TK, et al. Nivolumab plus Ipilimumab versus Sunitinib in Advanced Renal-Cell Carcinoma. N Engl J Med. 2018;378(14):1277-90.
4. Motzer RJ, Robbins PB, Powles T, Albiges L, Haanen JB, Larkin J, et al. Avelumab plus axitinib versus sunitinib in advanced renal cell carcinoma: biomarker analysis of the phase 3 JAVELIN Renal 101 trial. Nat Med. 2020;26(11):1733-41.
5. McDermott DF, Huseni MA, Atkins MB, Motzer RJ, Rini BI, Escudier B, et al. Clinical activity and molecular correlates of response to atezolizumab alone or in combination with bevacizumab versus sunitinib in renal cell carcinoma. Nat Med. 2018;24(6):749-57.
6. Motzer RJ, Banchereau R, Hamidi H, Powles T, McDermott D, Atkins MB, et al. Molecular Subsets in Renal Cancer Determine Outcome to Checkpoint and Angiogenesis Blockade. Cancer Cell. 2020;38(6):803-17 e4.
7. Chen Y-W, Beckermann K, Haake SM, Reddy A, Shyr Y, Atkins MB, et al. Optimal treatment by invoking biologic clusters in renal cell carcinoma (OPTIC RCC). Journal of Clinical Oncology. 2023;41(6_suppl):TPS742-TPS.
8. Tsakiroglou M, Evans A, Pirmohamed M. Leveraging transcriptomics for precision diagnosis: Lessons learned from cancer and sepsis. Front Genet. 2023;14:1100352.
9. Goh WWB, Wang W, Wong L. Why Batch Effects Matter in Omics Data, and How to Avoid Them. Trends Biotechnol. 2017;35(6):498-507.
10. Zhang Y, Parmigiani G, Johnson WE. ComBat-seq: batch effect adjustment for RNA-seq count data. NAR Genomics and Bioinformatics. 2020;2(3).
11. Cai Q, Christie A, Rajaram S, Zhou Q, Araj E, Chintalapati S, et al. Ontological analyses reveal clinically-significant clear cell renal cell carcinoma subtypes with convergent evolutionary trajectories into an aggressive type. EBioMedicine. 2020;51:102526.
12. Dagogo-Jack I, Shaw AT. Tumour heterogeneity and resistance to cancer therapies. Nature reviews Clinical oncology. 2018;15(2):81-94.
13. Lawson DA, Kessenbrock K, Davis RT, Pervolarakis N, Werb Z. Tumour heterogeneity and metastasis at single-cell resolution. Nature cell biology. 2018;20(12):1349-60.
14. Ing N, Huang F, Conley A, You S, Ma Z, Klimov S, et al. A novel machine learning approach reveals latent vascular phenotypes predictive of renal cancer outcome. Sci Rep. 2017;7(1):13190.
15. Graham S, Vu QD, Raza SEA, Azam A, Tsang YW, Kwak JT, Rajpoot N. Hover-Net: Simultaneous segmentation and classification of nuclei in multi-tissue histology images. Med Image Anal. 2019;58:101563.
16. Schmauch B, Romagnoni A, Pronier E, Saillard C, Maille P, Calderaro J, et al. A deep learning model to predict RNA-Seq expression of tumours from whole slide images. Nat Commun. 2020;11(1):3877.
17. Mondol RK, Millar EKA, Graham PH, Browne L, Sowmya A, Meijering E. hist2RNA: An Efficient Deep Learning Architecture to Predict Gene Expression from Breast Cancer Histopathology Images. Cancers (Basel). 2023;15(9).

18.	Alsaafin A, Safarpoor A, Sikaroudi M, Hipp JD, Tizhoosh HR. Learning to predict RNA sequence expressions from whole slide images with applications for search and classification. Commun Biol. 2023;6(1):304.
19.	Wang Y, Kartasalo K, Weitz P, Acs B, Valkonen M, Larsson C, et al. Predicting Molecular Phenotypes from Histopathology Images: A Transcriptome-Wide Expression-Morphology Analysis in Breast Cancer. Cancer Res. 2021;81(19):5115-26.
20.	Weitz P, Wang Y, Kartasalo K, Egevad L, Lindberg J, Gronberg H, et al. Transcriptome-wide prediction of prostate cancer gene expression from histopathology images using co-expression-based convolutional neural networks. Bioinformatics. 2022;38(13):3462-9.
21.	Falk T, Mai D, Bensch R, Cicek O, Abdulkadir A, Marrakchi Y, et al. U-Net: deep learning for cell counting, detection, and morphometry. Nat Methods. 2019;16(1):67-70.
22.	He K, Zhang X, Ren S, Sun J, editors. Deep residual learning for image recognition. Proceedings of the IEEE conference on computer vision and pattern recognition; 2016.
23.	Ricketts CJ, De Cubas AA, Fan H, Smith CC, Lang M, Reznik E, et al. The Cancer Genome Atlas Comprehensive Molecular Characterization of Renal Cell Carcinoma. Cell Rep. 2018;23(1):313-26 e5.
24.	Acosta PH, Panwar V, Jarmale V, Christie A, Jasti J, Margulis V, et al. Intratumoral Resolution of Driver Gene Mutation Heterogeneity in Renal Cancer Using Deep Learning. Cancer Res. 2022;82(15):2792-806.
25.	Kapur P, Christie A, Rajaram S, Brugarolas J. What morphology can teach us about renal cell carcinoma clonal evolution. Kidney Cancer J. 2020;18(3):68-76.
26.	Ohe C, Yoshida T, Amin MB, Atsumi N, Ikeda J, Saiga K, et al. Development and validation of a vascularity-based architectural classification for clear cell renal cell carcinoma: correlation with conventional pathological prognostic factors, gene expression patterns, and clinical outcomes. Mod Pathol. 2022;35(6):816-24.
27.	Wang T, Lu R, Kapur P, Jaiswal BS, Hannan R, Zhang Z, et al. An Empirical Approach Leveraging Tumorgrafts to Dissect the Tumor Microenvironment in Renal Cell Carcinoma Identifies Missing Link to Prognostic Inflammatory Factors. Cancer Discov. 2018;8(9):1142-55.
28.	Brugarolas J, Rajaram S, Christie A, Kapur P. The Evolution of Angiogenic and Inflamed Tumors: The Renal Cancer Paradigm. Cancer Cell. 2020;38(6):771-3.
29.	Shaban M, Raza SEA, Hassan M, Jamshed A, Mushtaq S, Loya A, et al. A digital score of tumour-associated stroma infiltrating lymphocytes predicts survival in head and neck squamous cell carcinoma. J Pathol. 2022;256(2):174-85.
30.	Barrera C, Corredor G, Viswanathan VS, Ding R, Toro P, Fu P, et al. Deep computational image analysis of immune cell niches reveals treatment-specific outcome associations in lung cancer. NPJ Precis Oncol. 2023;7(1):52.
31.	Harder N, Schonmeyer R, Nekolla K, Meier A, Brieu N, Vanegas C, et al. Automatic discovery of image-based signatures for ipilimumab response prediction in malignant melanoma. Sci Rep. 2019;9(1):7449.
32.	Palmer AC, Sorger PK. Combination Cancer Therapy Can Confer Benefit via Patient-to-Patient Variability without Drug Additivity or Synergy. Cell. 2017;171(7):1678-91 e13.
33.	Hanahan D, Weinberg RA. Hallmarks of cancer: the next generation. Cell. 2011;144(5):646-74.
34.	Gu YF, Cohn S, Christie A, McKenzie T, Wolff N, Do QN, et al. Modeling Renal Cell Carcinoma in Mice: Bap1 and Pbrm1 Inactivation Drive Tumor Grade. Cancer Discov. 2017;7(8):900-17.
35.	Cerami E, Gao J, Dogrusoz U, Gross BE, Sumer SO, Aksoy BA, et al. The cBio cancer genomics portal: an open platform for exploring multidimensional cancer genomics data. Cancer Discov. 2012;2(5):401-4.
36.	Gao J, Aksoy BA, Dogrusoz U, Dresdner G, Gross B, Sumer SO, et al. Integrative analysis of complex cancer genomics and clinical profiles using the cBioPortal. Sci Signal. 2013;6(269):pl1.

37.     Kapur P, Zhong H, Araj E, Christie A, Cai Q, Kim D, et al. Predicting Oncologic Outcomes in Small Renal Tumors. Eur Urol Oncol. 2022;5(6):687-94.
38.     Kapur P, Pena-Llopis S, Christie A, Zhrebker L, Pavia-Jimenez A, Rathmell WK, et al. Effects on survival of BAP1 and PBRM1 mutations in sporadic clear-cell renal-cell carcinoma: a retrospective analysis with independent validation. Lancet Oncol. 2013;14(2):159-67.
39.     Raulerson CK, Villa EC, Mathews JA, Wakeland B, Xu Y, Gagan J, Cantarel BL. SCHOOL: Software for Clinical Health in Oncology for Omics Laboratories. J Pathol Inform. 2022;13:1.
40.     Bankhead P, Loughrey MB, Fernandez JA, Dombrowski Y, McArt DG, Dunne PD, et al. QuPath: Open source software for digital pathology image analysis. Sci Rep. 2017;7(1):16878.
41.     Marstal K, Berendsen F, Staring M, Klein S, editors. SimpleElastix: A user-friendly, multi-lingual library for medical image registration. Proceedings of the IEEE conference on computer vision and pattern recognition workshops; 2016.
42.     Tellez D, Balkenhol M, Otte-Holler I, van de Loo R, Vogels R, Bult P, et al. Whole-Slide Mitosis Detection in H&E Breast Histology Using PHH3 as a Reference to Train Distilled Stain-Invariant Convolutional Networks. IEEE Trans Med Imaging. 2018.